\documentclass[draft]{aa}
\newcommand{\etal}{{\it et al.}}
\newcommand{\msun}{\mbox{$M_{\odot}\;$}}
\newcommand{\mstar}{\mbox{$M_{\ast}\;$}}
\newcommand{\rsun}{\mbox{$R_{\odot}\;$}}
\newcommand{\rstar}{\mbox{$R_{\ast}\;$}}

\newcommand{\kms}{\mbox{\ km\ s$^{-1}$}}
\newcommand{\mdot}{\mbox{$\stackrel{.}{\textstyle M}$}}
\newcommand{\msunyr}{\mbox{M$_{\odot}$\thinspace yr$^{-1}\;$}}
\newcommand{\ltappeq}{\mathrel{\hbox{\rlap{\hbox{\lower4pt\hbox{$\sim$}}}\hbox{$<$}}}}
\newcommand{\apj}{{\sl ApJ}}
\newcommand{\mnras}{{\sl MNRAS}}

\newcommand{\aanda}{{\sl A\&A}}
\newcommand{\aandasupp}[1]{{\sl Astr. Astrophys. Supp. Sers.} {\bf #1}}
\newcommand{\pasj}[1]{{\sl Publ. Ast. Soc. Japan} {\bf #1}}

\begin{document}

\title{On the optical--infra-red continuum emission from equatorial discs of supergiant
B[e] stars}

\author{John M. Porter}
                                                            
\offprints{jmp}
\mail{jmp@astro.livjm.ac.uk}

\institute{Astrophysics Research Institute, Liverpool John Moores University, 
Twelve Quays House, Egerton Wharf, Birkenhead, CH41 1LD, United Kingdom}

\date{Received 22/10/01 / Accepted 19/11/02}

\titlerunning{On the disc emission of supergiant B[e] stars}
\authorrunning{John M. Porter}

\abstract{
Two models of the circumstellar disc around supergiant B[e]
stars are discussed: an equatorial wind model produced by wind
bi-stability, and a Keplerian viscous disc model.
Both models are successful in providing a site for dust formation once
they have cooled sufficiently.
However, the optical--infra-red continuum is calculated and
it is found that both models have significant trouble in accounting for
observations. 
In particular the optical--near-IR emission is accounted for, but the
dust emission is underestimated by at least an order of magnitude.
Variations in the structure of the models (the temperature variation
with radius, the density structure and the dust opacity) are investigated to
assess how (in)appropriate the standard models are for supergiant B[e]
star discs.
Changing the temperature structure, and making simple dust opacity
changes within the disc has little effect
on the resultant continuum emission.
By altering the density structure of the discs, the
continuum may be accounted for by both models:  
the equatorial wind model requires a very flat density profile which
is impossible to explain with any accelerating wind, and the viscous disc
model's density structure is required to fall off less steeply with
radius than would have been expected, although this may be explained
from consideration of viscous processes in the disc.
It is recognised that both theoretical interpretations have
difficulties and unsolved problems.
\keywords{Stars: emission-line, Be -- supergiants -- 
circumstellar matter -- individual: R126 (HD37974)}
}
\maketitle 

\section{Introduction}

Observations of supergiant B[e] stars (denoted sgB[e] stars hereafter
-- see Lamers \etal\ 1998 for a classification system) 
show several features: 
infra-red (IR) continuum
excesses and far-IR dust emission (e.g. see Zickgraf
1992); broad ($\sim 10^3$\kms) UV resonance
lines; in the optical there are strong Balmer lines (H$\alpha$ equivalent
widths of $>10^2$\AA), and
narrow (10s of \kms) permitted and forbidden line emission.
These stars have 
typical stellar parameters of $\mstar\sim 30$--70\msun, $\rstar \sim
30$--100\rsun, $T_{\rm eff}\sim 15,000$--30,000K (Zickgraf \etal,
1986). 

Zickgraf \etal\ (1985) proposed that the observations could be explained
with a multicomponent wind structure: a fast line-driven wind
(appropriate for a hot star) over the poles and a dense disc of
outflowing gas in the equatorial plane.
This geometry has received some corroboration from polarimetry (Zickgraf
\& Schulte-Ladbeck 1989, Oudmaijer \etal\ 1998).

Whilst there is a good physical explanation of the fast polar wind,
(Castor, Abbott \& Klein, 1975)
the formation mechanism (and indeed the structure) of the dense
equatorial flow is still largely unknown. 
An excellent attempt at explaining the equatorial flow has been made
by Lamers \& Pauldrach (1991 -- rotationally-induced wind
bi-stability), which has been recently advanced by Pelupessy, Lamers \& Vink
(2000) who suggest that wind compression due to its rotation (Bjorkman
\& Cassinelli 1993) may also play a part in the generation of the disc
(although wind 
compression was found by Owocki, Cranmer \& Gayley 1996 and also by Petrenz \& 
Puls 2000 to be strongly inhibited in rotating winds).
Zickgraf \etal\ (1996) fit a wind-like flow with a ``beta'' type
velocity law to observations of the UV metallic lines of R50 and Hen
S22 (interpreted as edge-on stars) and found acceptable fits with 
wind terminal velocities of 60--80\kms.

Oudmaijer \etal\ (1998) present a hydrodynamical model of the flow
around HD87643 in which a line-driven wind is driven partly by
luminosity from a Keplerian reprocessing disc.  
As well as producing a fast polar wind, their
model produces a region several scale-heights above the equatorial
plane which has intermediate velocities ($\sim 100$\kms), and hence
they point to this region being responsible for the permitted line
emission. However, they did not discuss the disc formation, being more
concerned with the intermediate and fast radiatively driven flows.

Given the current uncertainties in the sgB[e] stars' disc origin, 
a model for an equatorial wind produced by the bi-stability
mechanism, and also a model for a viscous Keplerian disc (similar to
that currently gaining acceptance in the classical-Be star community) are
directly compared. 
The aim of this study is to assess if either model
can be ruled out, or (at least) if one model can be strongly favoured
over the other as the more likely disc structure of sgB[e] stars.
As a first step, the optical--IR spectral energy distribution is
calculated and compared to observations
(the line emission is returned to in the discussion). This is not a
trivial task as a successful model must also provide a site for the
formation of dust.
In \S2 the density and velocities of both the equatorial wind model
and the viscous disc model are discussed, along with a heuristic
argument for the temperature profile. The possibility of the flow
forming dust is considered in \S3.
In \S4 the continuum emission is calculated and its variation with
relevant parameters presented. A discussion is given in \S5 and
conclusions in \S6.

\section{The models}

The main concern of this study is to examine if {\em either} model can 
account for the continuum observations of sgB[e] stars. 
Most 
previous work has examined aspects of wind models (for example
Bjorkman, 1998, presents results of a wind compression model)
whilst this is
the first time a viscous disc model has been applied to sgB[e] stars.
During the
course of the study, values for the stellar parameters, fast polar
wind etc. are taken from the Large Magellanic Cloud star R126: $\mstar
\approx 40\msun$, $\rstar \approx 70\rsun$, $T \approx 22,500$K
(Zickgraf \etal\ 1985),
the fast polar wind parameters vary widely in the literature: a
mass-loss rate of $\mdot_w =
10^{-6}$--$10^{-5}\msunyr$, and terminal velocity of $v_\infty \approx
1800\kms$ was calculated by Zickgraf \etal\ (1985), whilst Bjorkman
(1998) gives $\mdot_w = 4.6\times 10^{-5}\msunyr$ and
$v_\infty = 650\kms$ from Kudritzki \etal's (1989) ``cooking recipe''.
The adopted nomenclature used in this paper is that mass-loss
rates will be quoted as a mass-flux multiplied by 4$\pi$ i.e. the rate
is quoted as if the mass-loss was present over the whole star. This
should be particularly noted when considering the mass flux through
the disc.

\subsection{Density and velocity structure}

\subsubsection{Equatorial wind model}
The popular model for the equatorial density enhancement of sgB[e]
circumstellar media is one based on radiatively-driven
winds. Support for this geometry came in 1991 when the bi-stability
model was presented, in which the stellar wind parameters (mass-loss rate and
velocity) 
``jump'' over a band centered on the equator of the star as the
driving species and lines change at temperatures 
in the region of 25,000K
due to the changing conditions at the photosphere 
from a gravity-darkened fast rotating star (Lamers \& Pauldrach 1991,
Vink \etal, 1999, and Pelupessy \etal, 2000)
The model is characterised by a latitudinal varying mass flux rate
$\mdot(\theta)$ and wind speed $v_{r}(\theta)$. At the jump latitude
$\theta$ , $\mdot(\theta)$
increases and $v_{r}$ decreases producing a large increase in density 
$\rho(\theta) \propto \mdot(\theta)/ r^2 v_r$.
Within this high density, low velocity region (representing the ``disc''),
the radial velocity is $v_{r,d}$ and density is $\rho_{d}$. 
The model requires that the star is rotating at a substantial fraction
($> 50\%$) of its critical speed (where centrifugal forces balance
gravity). The azimuthal velocity of the wind in the disc $v_{\phi,d}$
decreases with radius from the star due to angular momentum
conservation and so is not very important dynamically in the disc.
The disc thickness is determined by the value of $\theta$ 
where the bi-stability jump occurs, and so depends on the rotation
rate and how close the effective temperature is to the jump temperature. 

Within the disc, the density and radial and azimuthal velocities
follow the form 
\begin{eqnarray}
\rho_d & = & \frac{\mdot_d}{4\pi r^2 v_{r,d}} \\
v_{r,d} & = & w(r) v_{\infty,d} \\
v_{\phi,d} & = & v_{\phi 0} \left( \frac{r}{\rstar} \right)^{-1}
\end{eqnarray}
where
$v_{\phi 0}$ is the rotational speed at the stellar surface 
and again note the disc mass-loss rate is quoted as if this
mass loss applied over the whole star -- the actual mass-loss rate through
the disc is $\mdot_{d}$ multiplied by the solid angle of the disc
divided by 4$\pi$.
The velocity law is quoted in the form above with
$w(r) = ( 1 - R_0/r)^\beta$ where $R_0 = (1 - w_i)^{1/\beta}$,
and $w_i$ is the fraction of the terminal velocity that the flow has
at the star.
It is expected that the radial velocity at the stellar surface 
(corresponding to $w_i$) will be less than the sound speed.
With terminal velocities $v_\infty \sim 100$\kms (Zickgraf
\etal\ 1985)
and the sound speed $\sim 10$--$20\kms$ for B stars
this corresponds to $w_i \approx 0.1$.
Zickgraf (1992), estimates the disc opening angle to be
10--20$^\circ$ from the relative frequency of sgB[e] stars with
edge-on and polar characteristics, and this value is used hereafter.


\subsubsection{Viscous disc model}
The model of viscous outflowing discs has been discussed by Lee \etal\
(1991) and Okazaki (2001) and applied to classical Be star
observations by Porter (1999). 
In this model, the
angular momentum distribution within the disc is
determined by viscosity, in a similar way to the more familiar
accretion disc. However, in this case the angular momentum source for
viscous discs 
is the central star (which acts as an angular-momentum sink for
accretion discs). In common with the bi-stability mechanism, this
model {\em also} requires that the star is a
fast rotator -- indeed it insists that the equatorial region is
rotating at Keplerian speeds.
The weak link in the model for classical Be stars
(which will also feature in this application to sgB[e] stars) is that the
mechanism to inject angular momentum in to the disc at its inner
regions is still largely unknown -- often quoted candidates are pulsation
and magnetic fields.

The solution of the equations of mass continuity and angular momentum
conservation in the radial and azimuthal co-ordinates produce a disc
which has surface-density, radial and azimuthal velocity are
well represented in the inner $\sim 10^2$\rstar by power laws in
radius, (Okazaki 2001).

In particular the discs are close to Keplerian $v_\phi \propto
r^{-1/2}$, the radial velocity increases linearly with radius $v_r
\propto r$ (and is subsonic until large radii), and the
surface-density $\Sigma$ (the density integrated through the disc)
decreases as $\Sigma \propto r^{-2}$. The disc
flares -- the scale height $H$ of the disc increases with radius as $H
= c_s r^{1.5}/\sqrt{G\mstar}$ where $G$ is the gravitational
constant, \mstar is the stellar mass and $c_s$ is the sound
speed. 
The viscous disc
density and velocities are
\begin{eqnarray}
\rho_d & = & \rho_{0,d} \left( \frac{r}{\rstar} \right)^{-3.5} {\rm
exp}\left(-\frac{z^2}{2H^2}\right) \\
v_{r,d} & = & v_{r0,d} \left( \frac{r}{\rstar} \right)\\
v_{\phi,d} & = & v_{\phi 0} \left( \frac{r}{\rstar} \right)^{-0.5} 
\end{eqnarray}
where $\rho_{0,d}$, $v_{r0,d}$ and $v_{\phi 0}$ are the
density, radial velocity, and azimuthal velocity at the star-disc
boundary ($r = \rstar$) respectively, and $z$ is the height above the
equatorial plane. 
Also, $v_{r0,d} \ltappeq 10^{-2}c_s$ and $v_{\phi 0}
\approx \sqrt{G\mstar/\rstar}$ 
In the outer parts of the disc, it becomes transonic, and angular
momentum conserving (changes to $v_{\phi,d} \sim r^{-1}$).

\subsection{Temperature profile}

The form of temperature profile for a disc is difficult to
determine without a full calculation. However a simple argument can be
applied to both equatorial winds and viscous discs which can provide an
estimate of the radial variation of temperature.
It is assumed that the disc will have a constant temperature with
radius until a cooling radius $r_c$, and beyond this the disc will cool.
In the limit that {\em no} radiation of any wavelength
impinges on the disc, it will cool adiabatically such that $T \propto
r^{-4/3}$. 
However, it is most likely that the disc will act as a reprocessing
disc: it will absorb radiation from the star, and re-emit at its local
equilibrium temperature $T_d$. Hence the total energy emitted by the
disc ($\propto T_d^4$) is proportional to the flux recieved, which in
turn is proportional to $r^{-3}$ ($r^{-2}$ from the dilution of the
radiation field, and an extra factor of $r^{-1}$ from the cosine term
produced
by the geometric reduction of the projected disc area). 
Hence these reprocessing disc have $T_d^4 \propto r^{-3}$ or
$T_d \propto r^{-3/4}$.
This argument is valid for both equatorial winds, and viscous discs.

The cooling will change the metallic species responsible for the
absorption and scattering of the equatorial wind and hence may lead
to a more complicated discussion of the wind driving. 
This
has already been implicitly taken into account by using the slowly
accelerating beta-velocity law fitted to the UV lines from Zickgraf
\etal\ (1996).

Cooling will effect the viscous disc:
once the temperature in the viscous disc varies, then the density and
velocity profiles are also affected. This is due to the torques
changing as the viscosity $\nu = \alpha c_s H$ alters ($\alpha$ is the
Shakura \& Sunyaev 1973 viscosity parameter). This is a
result of the sound speed $c_s$ and also the disc scale
height $H \propto c_s$ changing. To ensure mass and
angular momentum conservation, the density, radial velocity and scale
height change to 
\begin{eqnarray}
\rho & \propto & r^{-19/8} \\
v_r & \propto & r^{1/4} \\
H & \propto & r^{9/8}
\end{eqnarray}
i.e. their radial dependence becomes less severe, 
whilst the azimuthal velocity remains Keplerian
(Okazaki 2001, private communication).
Therefore at radii where the viscous disc is cooling ($r > r_c$)
eqn.7--9 replace eqn.4--6.

\section{Dust formation}
Any model which attempts to explain the properties of sgB[e] star
circumstellar matter must also provide a site for dust formation. To
produce dust, two criteria need to be fulfilled: the temperature of
the gas must be lower than the sublimation temperature of the dust
(around 1500K depending on the chemical composition of the dust), and
secondly, the number density of the species involved in the formation
of the dust needs to be above a critical value.

It is expected that sgB[e] stars' dust will be silicate based, as
carbon is depleted due to the CNO processing cycle in
massive stars, and indeed sgB[e] stars do show features due to
silicates in their spectra (e.g. Voors, 1999). 
Examination of the 8-13$\mu$m spectra of R126 presented in Roche
\etal\ (1993) shows that there are no strong emission features, which
suggests silicate dust is not present.
However, suppression of the silicate feature
may be achieved by having large grains ($a_d>10\mu$m),
or by not
having a stratified disc structure (leading to any optically thin
``atmosphere'' region) i.e. the emitting region is completely
optically thick (see Dullemond \etal, 2001 and Meeus \etal\ 2002).

Gail \& Sedlmyar (1988) showed that a critical number density can be
derived by comparing the timescale
for the chemical reactions responsible for dust formation $\tau_{\rm
ch}$ to the timescale for a parcel of gas to change in
density and temperature $\tau_{\rm exp}$ (i.e. an expansion
timescale). Grain growth will occur if $\tau_{\rm ch} < \tau_{\rm exp}$.
Gail \& Sedlmyar find that this occurs for any size of grain if the
number density of the growth species is larger than 
a minimum density
\begin{equation}
n_{\rm min} = \frac{v_r}{10^{-16} r v_{\rm rel}}
\end{equation}
where $v_{\rm rel}$ is the relative velocity between the growing grain
cluster and a molecule (set to the thermal velocity
here).
It is assumed that dust formation occurs if $\epsilon \rho(r)/(\mu m_p) >
n_{\rm min}$ (where the ratio of relevant species
to total number density is $\epsilon$ and $\mu\approx 1.25^{-1}$ is the mean
molecular weight for neutral gas).
This criterion may be compared to differing models of the disc
component around sgB[e] stars.

The value of $\epsilon$ is taken to be 
$\epsilon = 10^{-5}$ corresponding to 1/3 of Solar silicon abundance
appropriate for the LMC metallicity.
As an example calculation of 
$n_{\rm nim}$, the sound speed (approximating to $v_{\rm rel}$) 
has been calculated assuming that for both the
equatorial wind and the viscous disc models, cooling takes place beyond
$r_c = 1.5\rstar$, and that the 
temperature profile is $T\propto (r/r_c)^{-3/4}$ thereafter.
Finally the temperatures at $r=\rstar$ are taken to be 0.8$T_{\rm
eff}$ and 0.5$T_{\rm eff}$ for the equatorial wind and disc models
respectively. The results of this calculation are shown in fig.1.

\begin{figure}
\vspace*{8.5cm}
\includegraphics{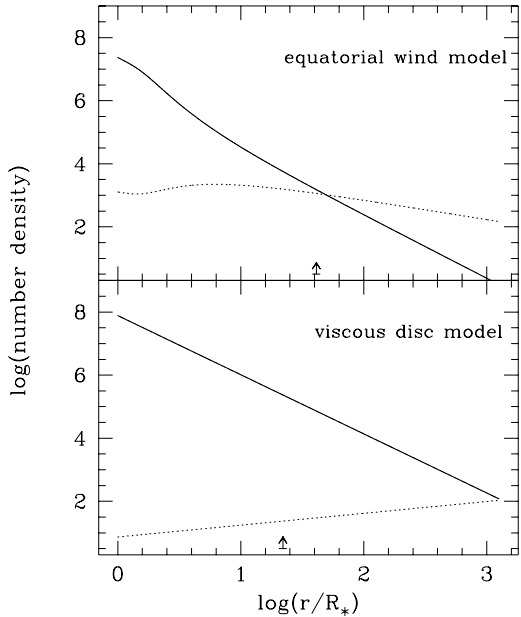}
\caption{Comparison of number density of carbon (solid line) with
$n_{\rm min}$ from eq.10 (dotted line): 
top panel, the equatorial wind model described in the text,
lower panel, the viscous disc model. Dust may
form in the region where the number density is larger than $n_{\rm
min}$ only if the flow is cool enough. The arrows indicate the radius
at which the temperature has fallen to 1500K.}
\end{figure}

\subsection{Equatorial wind model}
The equatorial wind is assumed to have the same parameters as
described in \S2.1.1:
($\beta = 4$ and $w_i = 0.1$) and a mass-loss rate of $\mdot_d =
10^{-5}$\msunyr (this rate has been chosen as it produces a dense
enough disc to acccount for the near-IR emission of the model star
R126 (see \S5).
Fig.1 shows the number density of this wind model as well as the
critical number density required to form dust.
It can be seen that dust can only form for $r <
50\rstar$. The arrow on fig.1 shows the radius where the temperature
falls to 1500K -- this occurs at a radius of 41\rstar. Therefore there
exists a region from $\sim 40$--50\rstar where the gas is dense enough
and cool enough for dust formation.

\subsection{Viscous disc model}
Fig.1 also shows the number density within a viscous disc
with $\rho_0 = 10^{-11}$g~cm$^{-3}$ (again chosen to reproduce the
near-IR emission, see \S5) and $v_{r0} = 0.3\alpha T_4
\sqrt{\rstar/\mstar}$\kms (where $\alpha = 0.1$ is the viscosity parameter,
$T_4$ is the disc temperature in $10^4$K and \rstar and \mstar are in
Solar units -- see eq. 6 of Porter, 1998). Again the cooling radius is
set to $r_c = 1.5\rstar$.
For $r\ltappeq 1250\rstar$ the number density in the disc is
high enough to enable dust to form. The radius where the temperature
falls to 1500K is 23\rstar, well within the region where dust may form.

\subsection{Dust formation conclusions}
Both the viscous disc model and the wind model have a region where
their number density is larger than the critical value for dust
formation, implying that both models can form dust, with the proviso
that the temperature must also be below the sublimation temperature
$T_{\rm dust}$.
The major difference between the models is for $n_{\rm min}$, which is
significantly lower for the viscous disc model as its radial
velocity is far lower than for the equatorial wind model.
Both models are successful in providing a site for dust
formation, and so both are still good candidates for the actual discs
od sgB[e] stars.
 
\section{Continuum emission}
The
continuum emission is now calculated from the density fields in the
models and compared to observations of R126. 
The model includes a fast polar wind (with a fixed mass-loss rate of
$10^{-5}\msunyr$), the disc models described in the last section, and
dust (where it exists) in the disc.
Although the fast polar wind is included, its contribution to the
final flux is negligible, and is only included for completeness.
In order to calculate the excess emission in the optical -- IR, the
prescription of Waters (1986) is used. 
The star is assumed to be pole-on. 

The viscous disc model requires that the density $\rho_{0,d}$, the dust
formation temperature $T_{\rm dust}$ and the cooling radius $r_c$
are specified to calculate the emission. The equatorial wind model
requires a mass-loss rate $\mdot_d$, a terminal velocity $v_{\infty,d}$, the
value of $\beta$ for the velocity law and a flow opening angle, as
well as the dust temperature and cooling radius: $\beta$ is fixed
at $\beta = 4$, the terminal velocity $v_\infty = 70$\kms, and the
opening angle to $20^\circ$ hereafter.

The number density
is calculated where the temperature falls below $T_{\rm dust}$ and
compared to the minimum number density for dust formation from eq.10. 
If dust is found to form, the optical depth through the disc at that
radii is 
\begin{equation}
\tau_{\rm dust} = \int_{\rm disc} 
\rho \kappa_{\rm dust} dz\nonumber
\end{equation}
where $\rho$ is the gas density, and $\kappa_{\rm dust}$ is the
opacity of the dust.  
The opacity used is an approximate fit to that calculated for a
distribution of grain sizes (e.g. see Wood \etal\ 2002).
A fit to Wood \etal's data for Solar metallicity in their
fig.2 from 0.3--100$\mu$m produces 
$\kappa_{\rm dust} = 120 (\lambda / 0.1\mu m)^{-0.6}\ {\rm cm}^2{\rm
g}^{-1}$.
The opacity used here is 1/3 of this, to take into account the lower
metallicity of the LMC. Therefore the dust opacity used here is 
\begin{equation}
\kappa_{\rm dust} = 40 \left(\frac{\lambda}{0.1\mu{\rm
m}}\right)^{-0.6}\ \ {\rm cm}^2{\rm g}^{-1}. 
\end{equation}
This simple expression for dust opacity is used so that the reason for
the failure of the standard viscous disc and equatorial wind models --
see below -- is clear.
In order that the dust is readily observed in the spectrum, it is
expected that the optical depth of the dust will be
$\tau_{\rm dust} > 1$.

The emitted flux is then integrated over the disc area,
which, for face-on discs, produces
\begin{equation}
F_\nu = \frac{1}{D^2} \int^{R_{\rm disc}}_{R_\ast} \pi B_\nu[T(r)] (1 - {\rm
e}^{-\tau_{\rm dust}} ) 2\pi r dr
\end{equation}
where $R_{\rm disc}$ is the outer edge of the disc, and $D$ is the
distance to the star and $B_\nu[T(r)]$ is the Planck function.

\section{Results}

The UBVRIJHKLMN photometry for R126 is taken from Zickgraf \etal's
(1985) compilation, and 
is supplemented by 12$\mu$m and 25$\mu$m IRAS points from Schwering
(1989). 
The reddening to the object was taken to be E(B-V)~=~0.25 from
Zickgraf \etal\ (1985). The calculated emission from the model has
been reddened with the expression given in Howarth (1983) and
the model flux in the V band has been normalised
to the R126 observations for all models. 
The underlying photospheric continuum is a Kurucz 22000K, log$g=3.0$
ATLAS9 model. 

The proceedure used to fit the photometry involves a search over the
mass-loss rate $\mdot_d$, dust formation temperature and cooling
radius for the 
equatorial wind model. The viscous disc model has the density
normalization $\rho_0$, dust formation temperature and cooling radius
varied to find the best fit.
The results are summarised in Table 1: the various parameters for the
models are given, along with their $\chi^2$ evaluation of fit.
Best-fit values to the optical--near-IR data produces values of
$\mdot_d = 1.1\times 10^{-5}$\msunyr and
$\rho_0 = 10^{-10.75}$g~cm$^{-3}$ for both models respectively. The
equivalent mass loss rate for the viscous disc model is $7\times
10^{-7}$\msunyr: although it appears counterintuitive that this is less
than the polar (fast wind) mass loss rate, it is reconciled in that
the mechanism for mass ejection into the viscous disc is not the same
as over the poles (indeed the cause for mass injection in to the disc
it is still unknown). Hence the equatorial mass loss rate need not
bear any relation to the polar mass loss rate. 
  
\begin{figure}
\vspace*{8.3cm}
\includegraphics{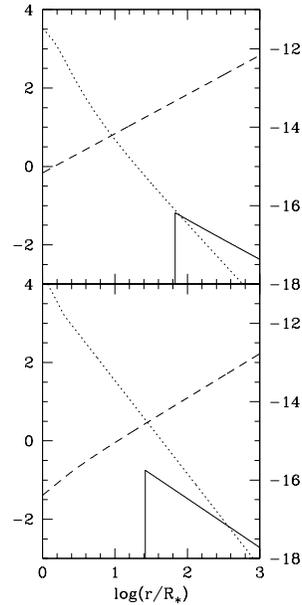}
\caption{Dust optical depths within the discs at a wavelength of
5$\mu$m for the two ``standard'' models. The solid line is the
log of the optical depth (referring to the left hand scale), the
dashed line is the log of the disc scale height $H$ in stellar radii
(again referring to the left hand scale), and the dotted line is the
log of the disc density in g~cm$^{-3}$ (the right hand scale). The top
panel is for the equatorial wind model, the bottom panel for the viscous
disc model.}
\end{figure}

There is a significant problem with the dust emission for both models
in that the optical depths where dust forms is not large enough to
generate the observed emission. To illustrate this, fig.2 shows the
dust optical depth $\tau_{\rm dust}$ (and density and disc
scale-height) for both models as a function of radius.
It can be clearly seen that the
optical depth is significantly less than unity at the dust formation
radius, and 
decreases with radius. Therefore it might be expected that the
continuum emission will not be able to reproduce the obserevational
data.
Fig.3 displays the full spectrum for both models. As can be seen the
optical to near-IR flux is reproduced, although longward of $\sim
2\mu$m the dust emission is too low by about an order of magnitude.

How may this be remedied? Or are the two proposed models to be
rejected?
To produce the observed emission, the optical depth $\tau_{\rm dust}$
needs to be increased by one to two orders of magnitude. For these
models this requires a significant increase in density. A global
increase in the density (corresponding to increases in $\mdot_d$ and
$\rho_0$) would produce too much emission in the optical--near-IR
range, and so may be ruled out. 
Increasing the opening angle for the equatorial wind model (or
decreasing the terminal velocity) will also
produce the same effect of increasing the emission at all wavelengths,
and not simply the dust emission.
To investigate how dependent the models are to the input assumptions,
several changes are made in turn to the standard models:
the temperature power law indices are made free parameters, the dust
opacity is set to that appropriate to Solar metallicity, and finally,
the density power law indices are made free parameters.
 
\begin{figure}
\vspace*{7.3cm}
\includegraphics{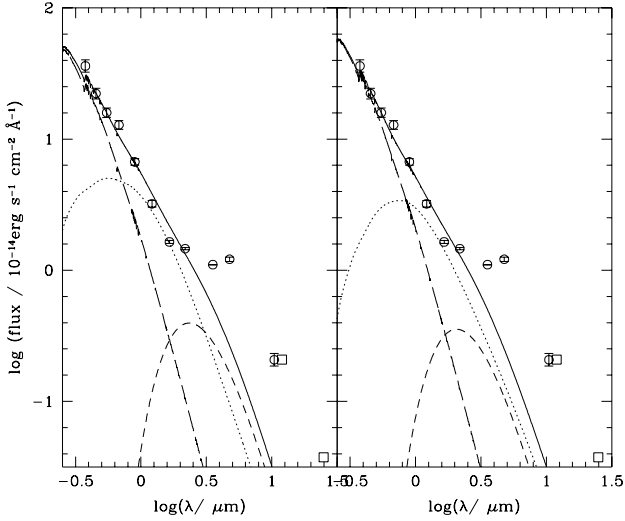}
\caption{Continuum emission from the star (R126) and the two best-fit models.
The total model flux (solid line) consists of Kurucz model atmosphere (long
dashed line), free-free and free-bound emission (dotted line) and dust
emission (short-dashed line) for the ``standard'' models. 
The model has been normalised to the R126 data of Zickgraf
\etal\ (1985) in the V band. 
The left panel is for the equatorial wind
model, and the right panel is for the viscous disc model.
Model parameters are given in Table 1.}
\end{figure}

\begin{table*}
\caption{The parameters for the best-fit models. The first two models
are the standard models with $m$ fixed at 0.75. 
The following two models have the temperature power law index $m$ as a
free parameter in the search 
and with the inner flow parameters as above. 
Models 5 \& 6 have an enhanced dust opacity appropriate for Solar metallicity.
The final two models have a power law in density as stated -- see text.
}
\begin{tabular}{lrrrrlr}
Model        &  $m$ & $r_c/\rstar$ & $T_{\rm dust}$ (K) & $\kappa_{\rm
dust}(0.1\mu$m) & density & $\chi^2$ \\ \hline
1. Wind         & 0.75 & 2.3 & 1400 & 40 & $\mdot_w= 1.1\times 10^{-5}\msunyr$ & 407 \\
2. Viscous disc & 0.75 & 2.0 & 1600 & 40 & $\rho=10^{-10.75}(r/\rstar)^{-3.5}$g~cm$^{-3}$ & 403 \\
3. Wind         & 0.60 & 1.1 & 1400 & 40 & $\mdot_w= 1.1\times 10^{-5}\msunyr$ & 359 \\
4. Viscous disc & 0.50 & 2.1 & 1500 & 40 & $\rho=10^{-10.7} (r/\rstar)^{-3.5}$g~cm$^{-3}$ &397 \\
5. Wind         & 0.75 & 2.0 & 1300 & 120 & $\mdot_w= 1.1\times 10^{-5}\msunyr$ & 316 \\
6. Viscous disc & 0.75 & 2.9 & 1400 & 120 & $\rho=10^{-10.7} (r/\rstar)^{-3.5}$g~cm$^{-3}$ &313 \\
7. Wind         & 0.75 & 3.5 & 700 & 40 & $\rho=10^{-11.9} (r/\rstar)^{-1.7}$g~cm$^{-3}$ & 64 \\
8. Viscous disc & 0.75 & 6.5 & 800 & 40 & $\rho=10^{-11.0} (r/\rstar)^{-2.7}$g~cm$^{-3}$ & 61 \\
\end{tabular}
\end{table*}

\subsection{The temperature profile}
It is possible that the temperature profile that has been used is not
correct, although the disc {\em must} cool in order for it to form
dust. 
To assess whether the observations may be fitted using a different
temperature profile, beyond the cooling radius, the power law $T_d
\propto r^{-m}$ has been varied using different values of $m$, from
0.4 to 1.5 and the best fit models calculated.
The best-fit parameters are shown in Table 1.
The temperature-power law index of the best models is $m = 0.6$ for the
equatorial wind and $m = 0.5$ for the viscous disc.

In the case of the equatorial wind the best-fit dust temperature is identical
to the standard wind model (1400K), and the cooling radius is
slightly smaller. However, the fit is again characterised by a significant
underestimate of the dust emission indicating that variation of the
temperature power law produces a negligible effect on the resultant
spectrum. 
The viscous disc produces a slightly cooler dust formation
temperature (1500K) than in the standard case (1600K).
However, as
in the wind case, the fit is barely improved with significant
underestimate of the dust emission.

Both of these models are presented in fig.4. It appears that the
failure of the standard models to reproduce the observations is not
dominated by the temperature profile in either model.

\begin{figure}
\vspace*{7.3cm}
\includegraphics{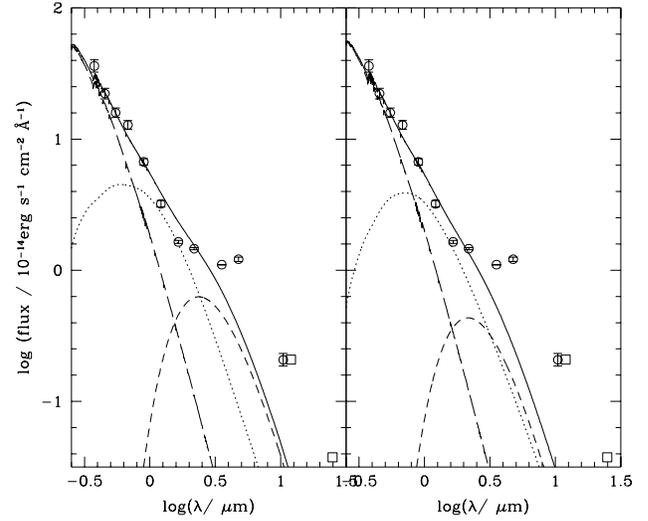}
\caption{As fig.3 except that the temperature power law index is a free
parameter. 
The left panel is for the equatorial wind
model with $T\propto r^{-0.6}$, and the right panel is for the
viscous disc model $T\propto r^{-0.5}$.
Further model parameters are given in Table 1.
Comparison with fig.3 illustraltes that the effect of allowing the
temperature power law to change is negligible.} 
\end{figure}

\subsection{The dust opacity}
Whilst so far the opacity from Wood \etal\ (2002) has been used, corrected
to take account of the lower metallicity of the LMC, here the
normalization for Solar metallicity is used (i.e. the LMC star R126 is
assumed to have Solar metallicity). 
This illustrates the
effect of increasing the opacity, whilst keeping the form of the
density field the same as in the standard cases.

The dust opacity used here is 
$\kappa_{\rm dust} = 120 (\lambda / 0.1\mu m)^{-0.6}\ {\rm cm}^2{\rm
g}^{-1}$, and the best-fitting models are shown in fig.5 (with
parameters in Table 1 : models 5 and 6).
Formally, from the $\chi^2$ values in Table 1, it can be seen that
increasing the dust opacity decreases the residuals. However, the best
fitting models (fig.5) are still woefully inadequate in accounting for
the dust emission.

Further increase in the dust opacity should provide a high enough optical
depth to account for observations. However, this would require an
exceptionally large metallicity for this star which is physically very
difficult 
to justify at this stage of evolution of the star, and therefore this
possibility is disregarded in favour of the dynamical arguments given later.

\subsection{The density profile}
The standard density profiles for the wind model (eq.1) is now changed
to a pure power law form: $\rho/\rho_0 = (r/\rstar)^{-n}$ (where $\rho_0$
is a normalization density at $r = \rstar$). 
The sole difference between the two models now is (i) that the equatorial
wind model 
has a disc with a constant opening angle (20$^\circ$), whereas the
viscous disc flares, and (ii) that the density falls exponentially
with height away from the equatorial plane for the viscous disc.
Only the temperature power
law index is fixed (with $m=0.75$), and the best-fit model search is
conducted over dust 
temperature $T_{\rm dust}$, density power law index $n$, density
normalization $\rho_0$, and cooling radius $r_c$.

The best-fitting equatorial wind model is shown in fig.6, and provides
an excellent description of the continuum flux of R126. The parameters
of this fit are $n = 1.7$ and $\rho_0 =
10^{-11.9}$g~cm$^{-3}$, cooling radius of $r_c = 3.5\rstar$ and dust
temperature of $T_{\rm dust} = 700$K. 
A literal interpretation of this best fit solution would predict that
the radial velocity actually decreases with radius (from comparison with eq.1),
which indicates that the simple wind model is an erroneous description of
the actual disc around R126.

The best fiting viscous disc model is also shown in fig.6 and also
provides a good fit to the data.
The density of this model is $\rho_0 = 10^{-11.0}$g~cm$^{-3}$ 
(note this is significantly higher than the best fitting equatorial
wind model), and the
density power law index is $n = 2.7$, the cooling radius is
$r_c = 6.5$ and the dust
formation temperature is $T_{\rm dust} = 800$K. 

For both the equatorial wind, and the viscous disc
the $\chi^2(n, \rho_0, r_c, T_{\rm dust})$ values have been inspected for
different models around this minimum, in order to 
estimated the variation in the parameters. 
A value of $\chi^2$ is obtained which is twice the best fitting
model if parameter
uncertainties are $\delta n \sim \pm 0.1$, $\delta \rho_0 \sim \pm 0.1$
$\delta T_{\rm dust} \sim \pm 100K$, and $\delta r_c \sim \pm
3$. Clearly the least constrained parameter is the cooling radius.

\begin{figure}
\vspace*{7.3cm}
\includegraphics{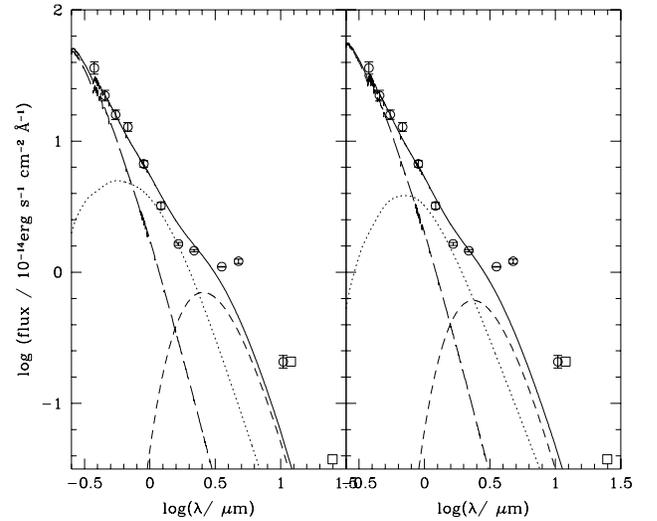}
\caption{As fig.3 except that the dust opacity normalization is appropriate
for Solar metallicity: $\kappa_{\rm dust} = 120
(\lambda/0.1\mu$m$)^{-0.6}$g~cm$^{-3}$.
Further model parameters are given in Table 1.
The left panel is for the equatorial wind
model and the right panel is for the viscous disc model.}
\end{figure}

\begin{figure}
\vspace*{7.3cm}
\includegraphics{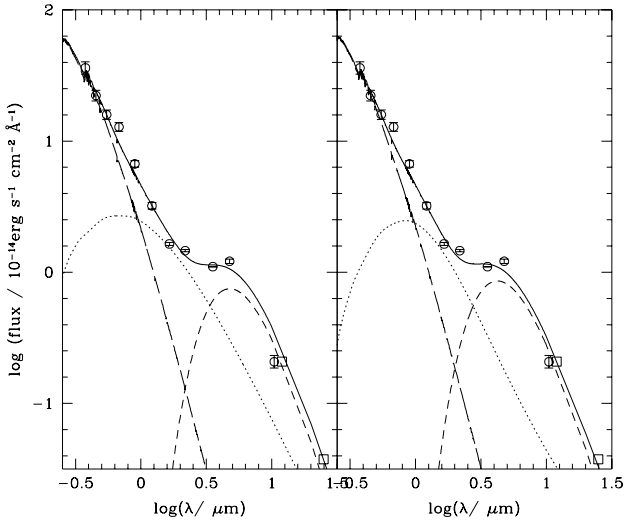}
\caption{As fig.3 except that the density power law is a free
parameter. 
The left panel is for the equatorial wind
model with $\rho = 10^{-11.9} (r/\rstar)^{-1.7}$g~cm$^{-3}$,
and the right panel is for the viscous disc model 
$\rho = 10^{-11.0} (r/\rstar)^{-2.7}$g~cm$^{-3}$. Other model
parameters are given in Table 1.}
\end{figure}

\section{Discussion}

It appears from the results that both the ``standard'' equatorial wind
model and 
the viscous disc model have trouble in generating the continuum
emission for sgB[e] stars. To do so, the density field has had to be
changed somewhat from initial theoretical expectations.
The deviations from the standard density structures of the models may
indeed highlight some aspects of the two models which have not been
considered here. The equatorial wind may be clumped at large radii,
hence producing parts of the disc which have a large dust optical
depth. However, the total emission may not necessarily increase due to 
the subsequent reduction in effective emitting area (i.e. if the gas
is clumped into dense regions, the ``filling factor'' of that dense gas
degreases).
The low power law index for the best fitting viscous disc model 
($n = 2.7$ compared to $n = 3.5$ for the standard model) might
indicate an extra dependence of the viscosity (and perhaps the
$\alpha$ parameterization) on the surface density, or the disc
temperature which has not been investigated here.

Assuming that either of these models can account for all the continuum
emission, 
can the two models both produce the observed heavy element line
emission from sgB[e] stars? The equatorial wind model was introduced
to do exactly that (e.g. see Zickgraf \etal\ 1996), and the line
emission comes from the equatorial wind flow itself. 
Therefore, this model would seem to provide a complete
description of the disc around a sgB[e] star.
However, there is a difficult issue for the equatorial wind
model: the electron scattering optical depth becomes larger than unity
at relatively small radii.
Whilst high
continuum optical depths are commonplace for (approximately) spherically
symmetric Wolf-Rayet winds, in this asymmetric case it may lead to
problems in 
driving the wind. Photons cannot physically penetrate into the disc
which leads to little or no radiative driving for the gas, and hence
the density profile may not actually allow the gas to be driven!
This might cast doubt on whether the equatorial wind model is correct
in principle. However, this may also highlight that the wind may not
have a velocity law typified by the ``beta'' parameterization.

The production of the heavy element line emission from the viscous
disc model is slightly more complicated. The dominant velocity
component in the disc is a rotational one, and therefore optically
thin emission lines should exhibit double-peaked profiles (similar to
those observed in classical Be stars). However, the model in Oudmaijer
\etal\ (1998), provides a separate site for line emission with a
dominant radial component of velocity: a disc wind. Here, the emission
from the reprocessing disc helps to drive a flow from
the upper and lower surfaces of the disc to relatively modest terminal
velocities $\sim 100$\kms. The resultant line emission may have a
double-peaked disc component, but will have a large radial component
imposed on it 
too. Currently, no attempt has been made to synthesize line profiles
from the work presented by Oudmaijer \etal\ An issue to resolve here is
whether there is enough flow generated in the disc-wind to account for
the lines observed in sgB[e] stars.

A weakness for the viscous
disc model is that the source of angular momentum required to supply
the disc is still unidentified (although candidates exist: pulsation
and/or magnetic fields). Until a
coherent theory of angular momentum transfer from the star to the disc
is produced, then the viscous disc model will only remain a promising
candidate. Note that
if the viscous disc model is the correct structure, then it is
possible that the way in which classical Be stars supply angular
momentum to their discs is the same for sgB[e] stars.
In this case then the sgB[e] stars will be invaluable as they have
a different envelope structure than classical Be stars, but
it might be likely that both supply angular momentum in the same
fashion.

From the results above, it is difficult to make a strong case
which will prefer one model over the other, or even that either model
is appropriate. Clearly, the discs of sgB[e] stars are not the simple
radiation driven flows, {\em nor} the simple viscous discs that have been
previously suggested.

\section{Conclusion}

Both of the models that have been presented in this study
are apparently incapable of reproducing the observational
results -- in their simplest form. Whilst both meet the
requirements for the formation of dust they are unable to
yield the necessary broad-band fluxes.

The first model is the enhanced equatorial flow from 
Lamers \& Pauldrach's bi-stability theory (1991); the other 
is the viscous disc model. The densities are alike, but the
velocity fields are disparate. Both competing paradigms 
expect that the central supergiant star is rotating at a
significant fraction of its break-up speed.

By relaxing several assumptions (regarding the temperature profile,
the dust opacity, and the density profile), the applicability of the
models have been tested. It is found that the models are unable to 
reproduce the observations unless the disc density field is
significantly altered, although then a good representation of the
observations can be obtained. 

There are possibilities for both models to produce the observed
permitted and
forbidden line emission: the equatorial wind model produces the
emission itself, and the viscous disc model produces a disc-wind from
it's surfaces which may produce the lines (although this remains
untested).

In both the equatorial wind model and the viscous disc model, there
is still work to be undertaken to understand the form of the density
profile necessary to account for observations. Clearly both models in
their simplest form are inadequate as a description of the discs of
sgB[e] stars.

\begin{acknowledgements}
JMP would like to thank Simon Clark, Atsuo Okazaki, Toby Moore
and Lee Howells for help in preparing and commenting on previous drafts
of this paper, and Jon Bjorkman for his useful input regarding dust emission.
JMP also thanks the anonymous referees for making useful
comments on the submitted versions.

\end{acknowledgements}

\end{document}